\documentstyle[12pt,epsf,aaspp4]{article}

\newcommand{\etal}{et~al.}
\newcommand{\bmbeta}{\mbox{\boldmath $\beta$}}
\newcommand{\bmtheta}{\mbox{\boldmath $\theta$}}

\lefthead{Kundi\'c, Cohen, Blandford \& Lubin}
\righthead{PG 1115+080 Lens Redshift}

\begin{document}

\title{Keck Spectroscopy of the Gravitational Lens System PG~1115+080:
       Redshifts of the Lensing Galaxies\altaffilmark{1}} 

\author{Tomislav Kundi\'c,\altaffilmark{2}
        Judith G. Cohen,\altaffilmark{3}
        Roger D. Blandford\altaffilmark{2} and
        Lori M. Lubin\altaffilmark{4}}

\altaffiltext{1}{Based on observations obtained at the W. M. Keck
Observatory, which is operated jointly by the California Institute of
Technology and the University of California}

\altaffiltext{2}{Theoretical Astrophysics, California Institute of
Technology, Mail Code 130-33, Pasadena, CA 91125; tomislav,
rdb@tapir.caltech.edu}

\altaffiltext{3}{Palomar Observatory, California Institute of
Technology, Mail Code 105-24, Pasadena, CA 91125;
jlc@astro.caltech.edu}

\altaffiltext{4}{The Observatories of the Carnegie Institution of
Washington, 813 Santa Barbara Street, Pasadena, CA 91101;
lml@ociw.edu}

\begin{abstract}

The quadruple system PG~1115+080 is the second gravitational lens with
a reported measurement of the Hubble constant. In addition to the
primary lens, three nearby galaxies are believed to contribute
significantly to the lensing potential. In this paper we report
accurate redshifts for all four galaxies and show that they belong to
a single group at $z_d = 0.311$. This group has very similar
properties to Hickson's compact groups of galaxies found at lower
redshifts. We briefly discuss implications for the existing lens
models and derive $H_0 = 52 \pm 14$ km s$^{-1}$ Mpc$^{-1}$.

\end{abstract}


\section{Introduction \label{intro.sec}}

Gravitational lensing provides a method of measuring Hubble's constant
at cosmological distances independently of the traditional distance
ladder. In a multiply-imaged source the difference in geometrical path
length and gravitational potential along each light ray results in a
time delay which can be measured if the source is variable. This time
delay is inversely proportional to $H_0$ with the constant of
proportionality that depends on the mass distribution in the lens
(Refsdal \markcite{REF64}1964).

Though simple in principle, Refsdal's method has been difficult to
implement, because of demanding observations required to determine the
time delays. In the double quasar 0957+561A,B (Walsh, Carswell, \&
Weymann \markcite{WCW79}1979), there has been a long controversy over
the correct value of the delay (Haarsma \etal\ \markcite{HAA97}1997
and references therein) that was only recently resolved by Kundi\'c
\etal\ (\markcite{KUN95}1995, \markcite{KUN96}1996). Combined with the
lens model of Grogin \& Narayan (\markcite{GN96}1996), the galaxy
velocity dispersion of Falco \etal\ \markcite{FAL97}(1997), and the
cluster mass model of Fischer \etal\ \markcite{FIS96}(1996), this time
delay yields $H_0 = 64 \pm 13$ km s$^{-1}$ Mpc$^{-1}$.

The second lens with a known time delay is the quadruple quasar
PG~1115+080 (Schechter \etal\ \markcite{SCH97}1997, Bar-Kana
\markcite{BAR97}1997). This system consists of four images of a $z_s =
1.722$ quasar (Weymann \etal\ \markcite{WEY80}1980) lensed by a
foreground group of galaxies at $z_d \sim 0.3$ (Henry \& Heasley
\markcite{HH86}1986; Angonin-Willaime, Hammer, \& Rigaut
\markcite{AHR93}1993). HST imaging of PG~1115+080 established the
relative positions of the four quasar images with an uncertainty of 5
mas and the centroid of the lensing galaxy with an uncertainty of 50
mas (Kristian \etal\ \markcite{KRI93}1993).  A schematic diagram of
the system is shown in Fig.~1 of Keeton \& Kochanek
\markcite{KK96}(1996, hereafter KK96). We adopt their galaxy
designations (based on Young \etal\ \markcite{YOU81}1981), in which
the primary lens is named G, and the nearby group members are named
G1, G2, and G3.

The goals of this paper are (1) to provide an accurate redshift for
the primary lensing galaxy and thus reduce the uncertainty in the
derived value of $H_0$, (2) to establish that G, G1, G2, and G3 belong
to the same group of galaxies, and (3) to estimate the velocity
dispersion of this group.

\section{Data Acquisition \label{data.sec}}

Optical spectra of PG~1115+080 were obtained on two nights, 1997
February 7 and 1997 March 2, with the Low Resolution Imaging
Spectrometer (Oke \etal\ \markcite{OKE95}1995) at the Keck II
10-m telescope. Table~1 lists the relevant observing parameters. The
first two exposures were taken with the 300 line/mm grating and the
0.7$\arcsec$ slit, and the other three with the 600 line/mm grating
and the 1.0$\arcsec$ slit. The resulting spectral resolution, as
measured from unresolved sky lines, was approximately 7 \AA\ in the
low-resolution spectra and 4.5 \AA\ in the high-resolution
spectra. The seeing was $\sim 0.7\arcsec$ in exposures 1 and 2, and
$\sim 1.2\arcsec$ in exposures 3--5.

\begin{figure}
\epsscale{0.9}
\plotone{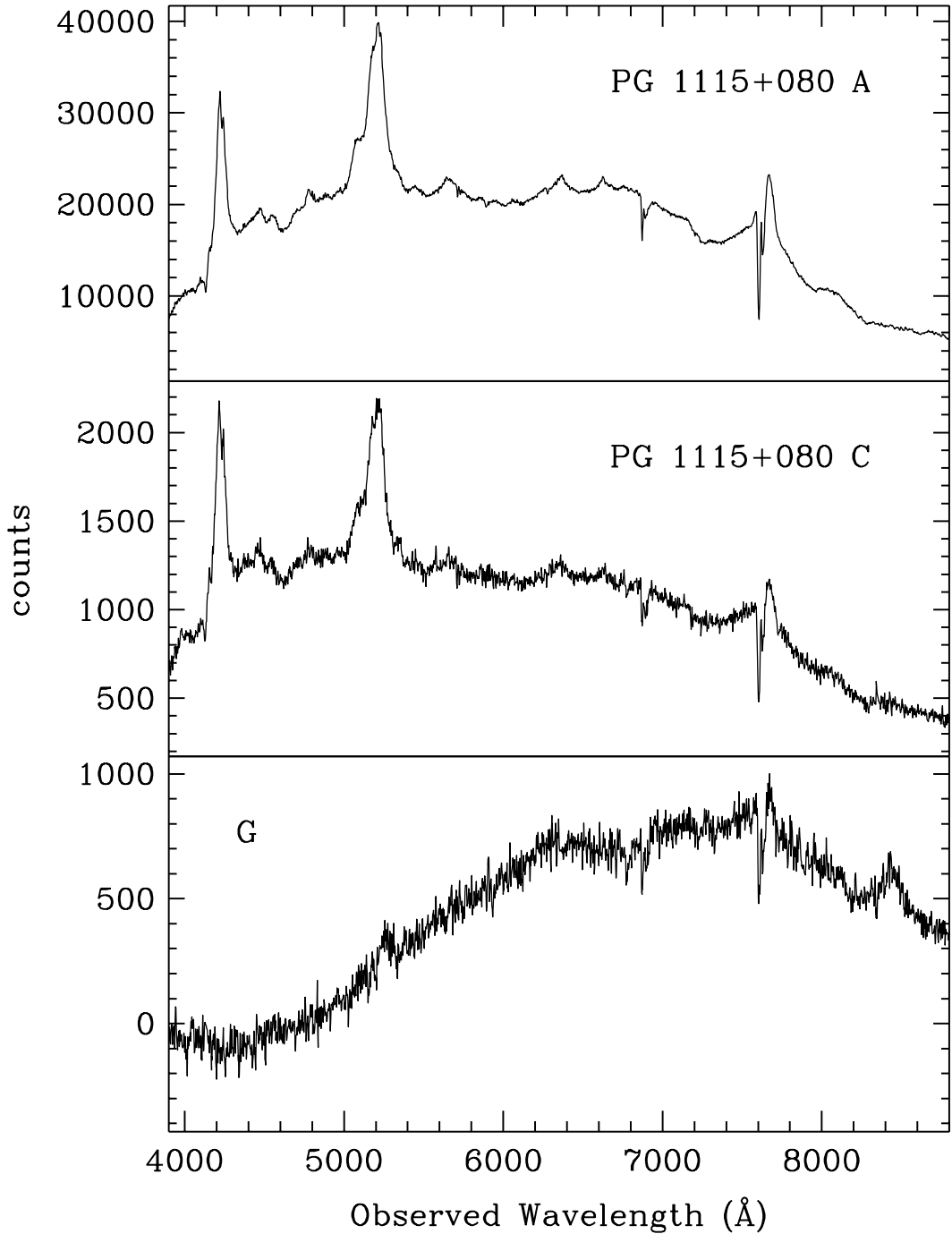}
\caption{The spectra of PG~1115+080 images A ({\em top}) and C ({\em
middle}), and the primary lensing galaxy G ({\em bottom}). Broad
emission line residuals in the G spectrum are caused by imperfect
quasar subtraction. The spectra have not been flux-calibrated.}
\label{ACG.fig}
\end{figure}

The spectra were reduced in a standard fashion using the ``longslit''
and ``apextract'' packages in IRAF\footnotemark\footnotetext{IRAF is
distributed by the National Optical Astronomy Observatories, which are
operated by the Association of Universities for Research in Astronomy,
Inc., under cooperative agreement with the National Science
Foundation.}. In the case of the main lensing galaxy G, we had to
subtract the contaminating quasar light from the galaxy spectrum.
Galaxy G is located approximately halfway between images A and C,
which are separated by about 2 arcseconds. In the $R$ band, where the
spectrograph is most sensitive, image A\footnotemark
\footnotetext{Image A consists of a close pair of images A1 and A2
which were unresolved in the original discovery paper (Weymann \etal\
\markcite{WEY80}1980)} is $\sim$4 magnitudes brighter than G, and
image C is $\sim2$ magnitudes brighter.  We thus proceeded by first
extracting the image A spectrum and then shifting the trace by 1 and 2
arcseconds to extract the spectra of galaxy G and image C. The
contribution of image A to the galaxy aperture was then subtracted
using an aperture on the opposite side of the image A center. The same
was repeated for image C.  The resulting galaxy spectrum is shown in
the bottom panel of Fig.~\ref{ACG.fig}, while the spectra of quasar
images A and C are shown in the top two panels. Note that no
correction for galaxy contamination was made to the quasar
fluxes. Some of the stronger galaxy absorption lines can thus be seen
in the image C spectrum.

For each spectrum the wavelength solution was derived using sky
emission lines identified in the atlas of Osterbrock \etal\
\markcite{OST96}(1996). After fitting a 4th order Legendre polynomial
to 10--20 strong, isolated lines, the rms residuals were typically
$\lesssim 0.3$ \AA. In order to minimize systematic errors, the sky
spectra were extracted on both sides of each galaxy spectrum,
independently calibrated and averaged for the final wavelength
solution. The shift between the two calibrations was in all cases
smaller than 0.5 \AA.

\begin{figure}
\epsscale{0.9}
\plotone{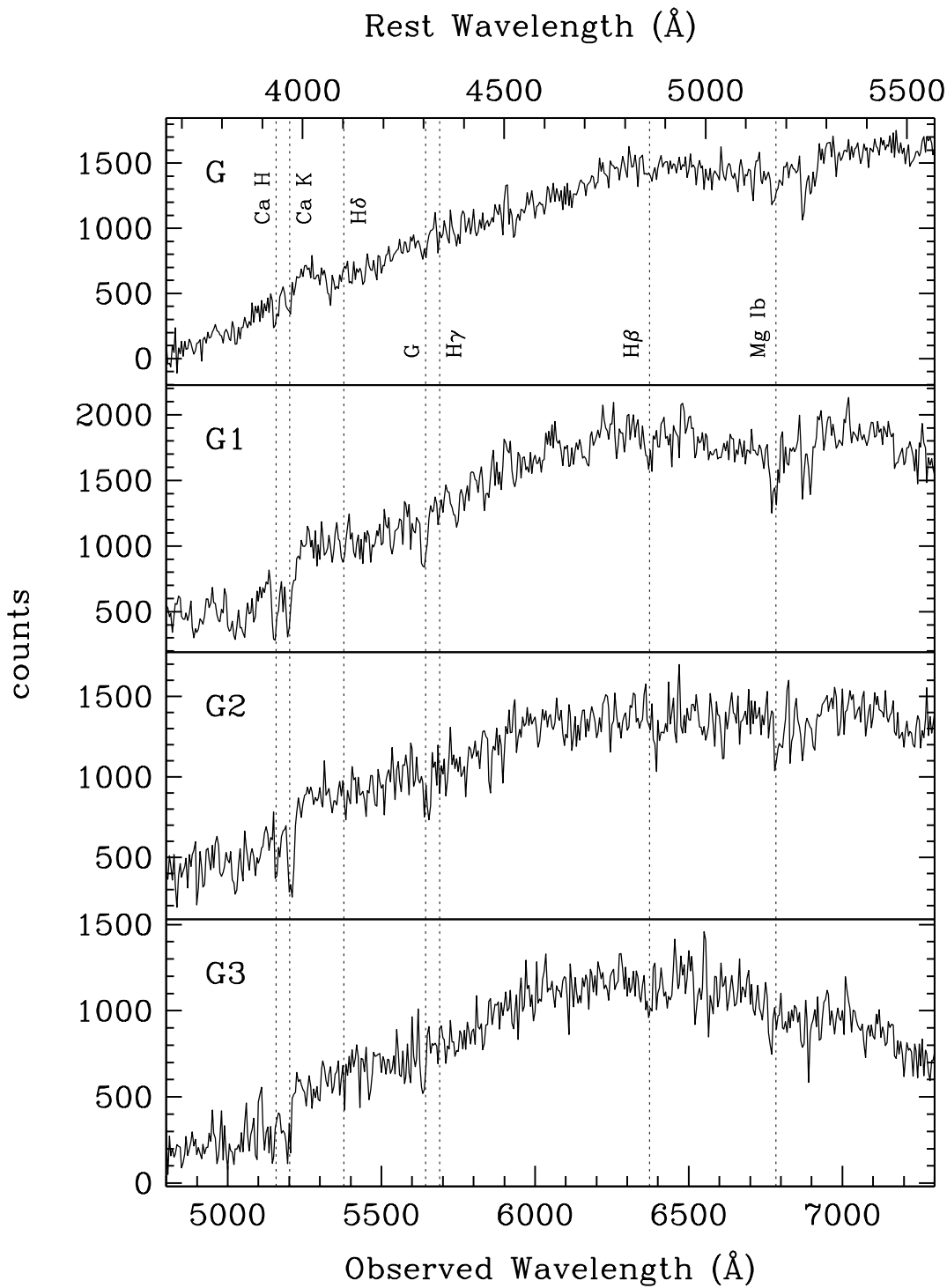}
\caption{The spectra of the primary lens G and group galaxies G1, G2,
and G3 ({\em top to bottom}) binned to 5 \AA\ resolution. Strong
spectral features are identified with dotted lines assuming the group
redshift of $z_d = 0.311$.  The same redshift is used for the rest
wavelength scale on the top of the figure. The redshift difference
between G2 and the other three galaxies is noticeable.}
\label{G123.fig}
\end{figure}

Approximate redshifts of the lensing galaxies were first estimated
from strong absorption features in the spectra, particularly the Ca II
H and K lines and the G band. These features, along with the Balmer
series lines and the MG Ib triplet, are marked with dotted lines in
Fig.~\ref{G123.fig}.  Accurate redshifts were then determined by
cross-correlating the galaxy spectra with stellar templates of Jones
\markcite{JON96}(1996) available on the AAS CD-ROM 7 (Leitherer \etal\
\markcite{LEI97}1997). Only G and K giants were used, since they
provide the closest match to the spectral characteristics of the early
type galaxies in the lensing group. The results of the
cross-correlation analysis are summarized in Table~2. The errors
listed in the table were calculated from the width of the
cross-correlation peak (Tonry \& Davis \markcite{TD79}1979) and an
estimate of the systematic error in wavelength calibration. Our
results are consistent with the previous measurement of the G1 and G2
redshifts by Henry \& Heasley \markcite{HH86}(1986), and marginally
consistent (at 3$\sigma$) with the G redshift reported by
Angonin-Willaime \etal\ \markcite{AHR93}(1993).

\section{Properties of the Lensing Group}

The spectra of G, G1, G2, and G3 clearly show that they belong to a
single group of galaxies at the redshift of $z_d = 0.311$
(Fig.~\ref{G123.fig}).  The rest frame line-of-sight velocity
dispersion of this group is $\sigma_v = c \sigma_z/(1 + z) = 270 \pm
70$ km s$^{-1}$, where the formal error includes only the uncertainty
in the individual galaxy redshifts. Because of the small number of
galaxies used to derive $\sigma_v$, the velocity dispersion of the
mass associated with the lensing group could be substantially larger
(Ramella \etal\ \markcite{RAM94}1994).

Properties of the lensing group in PG~1115+080 are very similar to
those of Hickson's compact groups (HCGs) of galaxies (Hickson
\markcite{HIC82}1982, Hickson \etal\ \markcite{HIC92}1992). The median
projected separation of galaxies in the lensing group is 35 $h^{-1}$
kpc (for $\Omega = 1$), as compared to $40^{+20}_{-10}$ $h^{-1}$ kpc
for HCGs (we quote the median with upper and lower quartiles). The
one-dimensional velocity dispersion of 270 km s$^{-1}$ is also within
the range of $200^{+100}_{-80}$ km s$^{-1}$ characteristic of
HCGs. The absolute magnitudes of the lensing galaxies are less
certain, but if we adopt $M_B = -19.1$ for the primary lens
(\markcite{KK96}KK96), the range of magnitudes in the lensing group,
$-20 \lesssim M_B \lesssim -19$, is compatible with $M_B({\rm HCG}) =
-19.5^{+0.8}_{-0.7}$.

The gravitational potential associated with the group is crucial for
the models of PG~1115+080, because it provides an independent source
of shear required to obtain a statistically acceptable fit to the
observables (\markcite{KK96}KK96; Keeton, Kochanek, \& Seljak
\markcite{KKS96}1996). As in the case of Q0957+561, however, the
presence of an extended perturber introduces a degeneracy into the
models (Falco, Gorenstein, \& Shapiro \markcite{FGS85}1985) that can
only be removed if the mass of the primary lens or the perturbing
group are independently measured. These estimates can be obtained from
the line-of-sight velocity dispersions of the galaxy and the
group. The former requires high signal-to-noise, moderate ($\sim 1$
\AA) resolution spectroscopy of the lensing galaxy (e.g. Falco \etal\
\markcite{FAL97}1997), a challenging observation because of the
proximity of bright quasar images. The velocity dispersion of the
group can be improved from our current estimate if additional galaxies
associated with the group are found in the vicinity of the lens
(Ramella \etal\ \markcite{RAM94}1994). A deeper spectroscopic survey
of the field would thus be highly desirable.

\section{Implications for the Hubble Constant}

In the expression for the gravitational lens time delay, it is common
to separate the lens model dependence from the cosmological scale
factor (e.g. Blandford \& Narayan \markcite{BN92}1992): 
\begin{equation}
\tau(\bmtheta) = \frac{(1 + z_d)}{c} \frac{D_d D_s}{D_{ds}} \left[
\frac{(\bmtheta - \bmbeta)^2}{2} - \Psi(\bmtheta) \right]
\quad ,
\end{equation}
where $\bmtheta$ and $\bmbeta$ are the image and source positions;
$\Psi(\bmtheta)$ is the scaled surface potential; and $D_d$, $D_s$,
and $D_{ds}$ are the angular diameter distances observer--lens,
observer--source, and lens--source. In a given lens model specified by
$\Psi(\bmtheta)$, the value of the Hubble constant derived from the
differential time delays, $\tau(\bmtheta_i) - \tau(\bmtheta_j)$, is
inversely proportional to $K = (1 + z_d)/c \; (D_d D_s)/D_{ds}$. This
factor $K$ also depends on the source and lens redshifts, and the
choice of cosmological parameters $\Omega_0$ and $\Lambda$.

For the source redshift of $z_s = 1.722$ and the lens redshift of $z_d
= 0.311$, the cosmological scale factor $K$ takes the values of 31.3,
33.7, 32.6, and 32.7 $h^{-1}$ days arcsec$^{-2}$ when $(\Omega_0,
\Lambda)$ = $(1, 0)$, $(0.1, 0)$, $(0.4, 0.6)$, and $(0.2, 0.8)$
respectively.  In the $(\Omega_0, \Lambda)$ = $(1, 0)$ cosmology, the
lens model of KK96 then implies $H_0 = 52 \pm 14$ km s$^{-1}$
Mpc$^{-1}$. This value is approximately 3\% higher than the one quoted
by KK96 who use $z_d = 0.304$. It is worth noting that $K$ is much
less sensitive to $\Omega_0$ and $\Lambda$ than the angular diameter
distances involved, making the derived value of $H_0$ robust with
respect to the choice of the world model (Blandford \& Kochanek
\markcite{BK87}1987, Blandford \& Kundi\'c \markcite{BK97}1997). For
the same reason, the time delay method is not an effective way to
constrain $\Omega_0$ and $\Lambda$.

\section{Conclusions}

In this paper we demonstrate that the main lensing galaxy in the
gravitational lens system PG~1115+080 and its three neighbors belong
to a single group at $z = 0.311$.  With its velocity dispersion of
$270 \pm 70$ km s$^{-1}$ and median projected galaxy separation of 35
$h^{-1}$ kpc, this group is very similar to Hickson's compact groups
of galaxies discovered at lower redshifts.  The presence of the group
is important for the models of the system, because it provides an
additional source of shear required to explain the observed image
configuration.  Such a two--shear model has been constructed by
\markcite{KK96}KK96.  Using Bar-Kana's \markcite{BAR97}(1997) analysis
of Schechter \etal\ \markcite{SCH97}(1997) light curves, the KK96
model implies $H_0 = 52 \pm 14$ km s$^{-1}$ Mpc$^{-1}$ in an
Einstein-DeSitter universe.

\acknowledgments
We thank D. W. Hogg and S. Malhotra for helpful conversations. This 
work was supported by NSF grant AST~95--29170 and NASA grant NAG~5-3834. 

\clearpage

\begin{deluxetable}{ccccccccc}
\footnotesize
\tablecolumns{9}
\tablewidth{0pc}
\tablecaption{Observing Parameters}
\tablehead{
   \colhead{Exposure} & \colhead{Object} & \colhead{UT} &
   \colhead{UT} & \colhead{Airmass} & \colhead{PA} & 
   \colhead{Exposure} & \colhead{Grating} &
   \colhead{Wavelength} \nl
   \colhead{Number} & & \colhead{Date} & \colhead{Time} & &
   \colhead{($\arcdeg$E of N)} &
   \colhead{Time ($s$)} & & \colhead{Range (\AA)} 
}
\startdata
1 & G      & 1997 Feb 7 & 09:51 & 1.31 & 298 &  600 & 300/5000 & 3900-8800 \nl
2 & G      & 1997 Feb 7 & 10:04 & 1.26 & 298 &  600 & 300/5000 & 3900-8800 \nl
3 & G2, G3 & 1997 Mar 2 & 09:02 & 1.16 & 192 & 1200 & 600/5000 & 4800-7300 \nl
4 & G2, G3 & 1997 Mar 2 & 09:31 & 1.10 & 192 & 1200 & 600/5000 & 4800-7300 \nl
5 & G1, G3 & 1997 Mar 2 & 09:54 & 1.06 & 280 & 1200 & 600/5000 & 4800-7300 \nl
\enddata
\end{deluxetable}

\begin{deluxetable}{crrcc}
\tablecolumns{5}
\tablewidth{0pc}
\tablecaption{Lensing Galaxy Redshifts}
\tablehead{
   \colhead{Galaxy} & \colhead{$-\Delta \alpha$ ($\arcsec$)} &
   \colhead{$\Delta \delta$ ($\arcsec$)} & \colhead{Magnitude} &
   \colhead{Redshift}
}
\startdata
G  & $-0.4$\phn\phn &  $-1.3$\phn\phn & $R = 20.2$ & $0.3100 \pm 0.0005$ \nl
G1 & $20.1$\phn\phn & $-12.3$\phn\phn & $r = 19.0$ & $0.3099 \pm 0.0005$ \nl
G2 & $11.5$\phn\phn &  $-2.1$\phn\phn & $r = 20.0$ & $0.3120 \pm 0.0005$ \nl
G3 & $13.8$\phn\phn & $-13.5$\phn\phn & $r = 20.5$ & $0.3093 \pm 0.0005$ \nl
\enddata
\tablecomments{Galaxy positions are given relative to
image C, based on the data from Young \etal\ (1991) and Kristian
\etal\ (1993). The magnitude of G is taken from Christian, Crabtree,
\& Waddell (1987), while the other three magnitudes are adopted from
Young \etal\ (1991).}
\end{deluxetable}

\clearpage

\end{document}